\documentclass[a4paper,11pt]{article}
\usepackage[left=3.17cm,right=3.17cm,top=2.54cm,
headheight=0.5cm,headsep=0.54cm,bottom=2.54cm,footskip=0.79cm
]{geometry}

\usepackage{amsmath}
\usepackage[pdfstartview=FitH,bookmarks=false]{hyperref}

\begin{document}

\title{Deformation quantization for
coupled harmonic oscillators on a general noncommutative
space}

\author{Bingsheng Lin, Sicong Jing$^{\dag}$ and Taihua Heng\\
\textit{\small{Department of Modern Physics, University of Science
and Technology of China}}\\
\textit{\small{Hefei, Anhui 230026, China}}\\
$^{\dag}$sjing@ustc.edu.cn
}
\date{25 January 2007}

\maketitle

\begin{abstract}
\noindent 
Deformation quantization is a powerful tool to quantize
some classical systems especially in noncommutative space. In this
work we first show that for a class of special Hamiltonian one can
easily find relevant time evolution functions and Wigner functions,
which are intrinsic important quantities in the deformation
quantization theory. Then based on this observation we investigate a
two coupled harmonic oscillators system on the general
noncommutative phase space by requiring both spatial and momentum
coordinates do not commute each other. We derive all the Wigner
functions and the corresponding energy spectra for this system, and
consider several interesting special cases, which lead to some
significant results.
\end{abstract}

\section{Introduction}
Deformation quantization \cite{s1}, based on Wigner's
quasi-distribution function \cite{s2} and Weyl's correspondence
between quantum operators and ordinary $c$-number phase-space
functions \cite{s3}-\cite{s3b}, have be extensively studied (for
recent reviews, see e.g. Ref.~\cite{s4}-\cite{s7}).

In recent years, there has been much interest in the study of
physics in noncommutative space. The ideas of noncommutative
space-time and field theories defined on such a structure started
already in 1947 \cite{s8,s8a}. It came up again first in the 1980's,
when Connes formulated the mathematically rigorous framework of
noncommutative geometry \cite{s9,s9a}. A noncommutative space-time
first appeared in physics was in the string theory, namely in the
quantization of open string \cite{s10}, and the noncommutativity of
space-time also plays an important role in quantum
gravity \cite{s11}. Also in condensed matter physics the concept of
noncommutative space-time is applied, such as the integer quantum
Hall effect \cite{s12}. Since the noncommutativity between spatial
and time coordinates may lead to some problems with unitary and
causality, usually only spatial noncommutativity is considered. So
far quantum theory on the noncommutative space has been studied
extensively, and the main approach is based on the Weyl-Moyal
correspondence which amounts to replacing the usual product by a
$\ast$-product in the noncommutative space. Therefore, deformation
quantization has special significance in the study of physical
systems on the noncommutative space.

Some works on harmonic oscillators in the noncommutative space from
the point of view of deformation quantization have been
reported \cite{s13}. Considering there are many physical models based
on coupled harmonic oscillators \cite{s14}-\cite{s16b}, it is
interesting to investigate the coupled harmonic oscillators on the
noncommutative space. In Ref.~\cite{s17} two coupled harmonic
oscillators on noncommutative plane with space-space
noncommutativity were studied, and some results were obtained. In
the paper the authors changed the noncommutative problems into
problems in the usual commutative space with a well-known
coordinates transformation in phase space, and solved the problems
with normal quantum mechanics method. But there will be some
difficulties of diagonalizing the new Hamiltonian obtained by the
coordinates transformation in phase space, and the results such as
the Wigner functions are still beyond to the commutative phase
space, had not been transformed to those of the noncommutative
space. So in the present paper we will remain in noncommutative
space and use a new method of deformation quantization to solve this
problem.

First we show that for a class of Hamiltonian with special form one
can easily derive the relevant $\ast$-Exponential function which
determines the time evolution of the systems, and all the Wigner
functions may be obtained by a Fourier-Dirichlet expansion of the
$\ast$-Exponential function. When using this observation to the
coupled harmonic oscillators on a general noncommutative phase space
with both the spatial and momentum coordinates being noncommutative,
we derive the explicit form of the relevant $\ast$-Exponential
function, and then get all the Wigner functions and the
corresponding energy spectra of the two coupled harmonic oscillators
system on the noncommutative phase space.

This paper is organized as follow. In Sec.~\ref{sec2} we brief\/ly
review the fundamental concepts of deformation quantization of a
classical system on the noncommutative space. In Sec.~\ref{sec3} we
discuss a class of Hamiltonian with special form and present some
practical formulas. In Sec.~\ref{sec4} we study the properties of a
system consisting of two coupled harmonic oscillators on the
noncommutative phase space, derive its energy spectra and all the
Wigner functions. In Sec.~\ref{sec5} we discuss some special cases
of harmonic oscillators system based on the results in
Sec.~\ref{sec4}, and compare the results with those we have known in
the literature. The summary and concluding remarks are in
Sec.~\ref{sec6}~.

\section{Deformation quantization on noncommutative phase space}\label{sec2}
Consider a 4$D$ general noncommutative phase space, the coordinates
of position and momentum are denoted by
$\boldsymbol{x}$=$\{x_1,x_2\}$ and $\boldsymbol{p}$=$\{p_1,p_2\}$~,
and their corresponding quantum operators $\hat{\boldsymbol{x}}$ and
$\hat{\boldsymbol{p}}$ satisfy the following commutation
relations \cite{s18}
\begin{equation}\label{nc}
    [\hat{x}_{i},\hat{x}_{j}]=\mathrm{i}\epsilon_{ij}\mu~,
    \quad[\hat{p}_{i},\hat{p}_{j}]=\mathrm{i}\epsilon_{ij}\nu~,
    \quad[\hat{x}_{i},\hat{p}_{j}]=\mathrm{i}\delta_{ij}\hbar~,
\end{equation}
where $i,j=1,2$~, and $\boldsymbol{\epsilon}$ is the antisymmetric
matrix
\[\boldsymbol{\epsilon}=
\begin{pmatrix}
  ~0 & ~1\\
  -1 & ~0
\end{pmatrix}.
\]

In the deformation quantization theory of a classical system in the
noncommutative space, one treats $(\boldsymbol{x}, \boldsymbol{p})$
and their functions as classical quantities, but replaces the
ordinary product between these functions by the following
generalized $\ast$--product \cite{s6}
\begin{eqnarray}\label{gstar}
    \ast&=&\ast_{\hbar}\ast_{\mu}\ast_{\nu}\nonumber\\
    &=&\exp \left\{\frac{\mathrm{i}\hbar}{2}\Big(\overleftarrow{\partial}\!_{x_{i}}
    \overrightarrow{\partial}\!_{p_{i}}-\overleftarrow{\partial}\!_{p_{i}}\overrightarrow{\partial}\!_{x_{i}}\Big)
    +\frac{\mathrm{i}\mu}{2}\epsilon_{ij}\overleftarrow{\partial}\!_{x_{i}}\overrightarrow{\partial}\!_{x_{j}}
    +\frac{\mathrm{i}\nu}{2}\epsilon_{ij}\overleftarrow{\partial}\!_{p_{i}}\overrightarrow{\partial}\!_{p_{j}}\right\},
\end{eqnarray}
here we have used the Einstein summation convention and also in the
latter part of this article without additional indication. The
variables $x_i$, $p_i$ on the noncommutative phase space satisfy the
following commutation relations similar to (\ref{nc})
\begin{equation}\label{mbnc}
    [x_{i},x_{j}]_{\ast}=\mathrm{i}\epsilon_{ij}\mu~,
    \quad[p_{i},p_{j}]_{\ast}=\mathrm{i}\epsilon_{ij}\nu~,
    \quad[x_{i},p_{j}]_{\ast}=\mathrm{i}\delta_{ij}\hbar~,
\end{equation}
where the Moyal bracket is defined as $[f,g]_{\ast}=f\ast g-g\ast
f$~.

The time evolution function for a time-independent Hamiltonian $H$
of the system is denoted by the $\ast$-Exponential function, which
is the solution of the following equation
\begin{eqnarray}\label{time}
\lefteqn{\mathrm{i}\hbar\frac{\mathrm{d}}{\mathrm{d}t}\mathrm{Exp}\left(\frac{Ht}{\mathrm{i}\hbar}\right)
    =H(\boldsymbol{x},\boldsymbol{p})\ast
    \mathrm{Exp}\left(\frac{Ht}{\mathrm{i}\hbar}\right)}\nonumber\\
    &&~~~~~~~~=H\Big(x_{i}+\frac{\mathrm{i}\hbar}{2}\overrightarrow{\partial}\!_{p_i}
    +\frac{\mathrm{i}\mu}{2}\epsilon_{ik}\!\overrightarrow{\partial}\!_{x_k}~,
    ~p_{j}-\frac{\mathrm{i}\hbar}{2}\overrightarrow{\partial}\!_{x_j}
    +\frac{\mathrm{i}\nu}{2}\epsilon_{jl}\!\overrightarrow{\partial}\!_{p_l}\Big)
    \mathrm{Exp}\left(\frac{Ht}{\mathrm{i}\hbar}\right).~~~~
\end{eqnarray}
where $i,j,k,l=1,2$. Eq. (\ref{time}) corresponds to the
time-dependent Schr\"odinger equation, and the $\ast$-Exponential
function can be expressed by \cite{s6}
\begin{equation}\label{expd}
    \mathrm{Exp}\left(\frac{Ht}{\mathrm{i}\hbar}\right)
    :=\sum_{n=0}^{\infty}\frac{1}{n!}\left(\frac{t}{\mathrm{i}\hbar}\right)^{n}(H\ast)^{n},
\end{equation}
where
\[
(H\ast)^{n}=\underbrace{H\ast H\ast\cdots\ast H}_{n\
\mathrm{times}}~.
\]
The generalized $\ast$--genvalue equation is
\begin{equation}\label{eigen}
    H\ast\mathcal{W}_{n}=\mathcal{W}_{n}\ast
    H=E_{n}\mathcal{W}_{n}~,
\end{equation}
where $\mathcal{W}$ is the Wigner function and $E$ is the
corresponding energy eigenvalue of the system. Eq. (\ref{eigen})
corresponds to the time-independent Schr\"odinger equation. The
relation between the Wigner functions and the $\ast$-Exponential
function is:
\begin{equation}\label{we}
    \mathrm{Exp}\left(\frac{Ht}{\mathrm{i}\hbar}\right)
    =\sum_{n=0}^{\infty}e^{-\mathrm{i}E_{n}t/\hbar}\mathcal{W}_{n}~,
\end{equation}
this expression is also called Fourier-Dirichlet expansion for the
time-evolution function.

\section{A class of Hamiltonian with special form}\label{sec3}
Generally in order to obtain the $\ast$-Exponential function or the
Wigner functions for some systems, one may try to solve the
differential equation (\ref{time}) or (\ref{eigen}) which contains
five or four variables. In many cases it is very difficult, but when
the Hamiltonian is some special form, the problem can be
simplified.

Now let us consider a Hamiltonian which can be written as a sum of
two perfect square parts
\begin{equation}\label{hsq}
H=\left(\boldsymbol{a}\cdot\boldsymbol{x}+\boldsymbol{b}\cdot\boldsymbol{p}\right)^2
+\left(\boldsymbol{c}\cdot\boldsymbol{x}+\boldsymbol{d}\cdot\boldsymbol{p}\right)^2~,
\end{equation}
where $\boldsymbol{a}$=$\{a_1,a_2\}$,
$\boldsymbol{b}$=$\{b_1,b_2\}$, $\boldsymbol{c}$=$\{c_1,c_2\}$ and
$\boldsymbol{d}$=$\{d_1,d_2\}$, and the coefficients $a_i$, $b_i$,
$c_i$ and $d_i$ are arbitrary real constants. The time evolution
equation (\ref{time}) with the Hamiltonian~(\ref{hsq}) will become
\begin{eqnarray}\label{ph}
\lefteqn{\mathrm{i}\hbar\frac{\mathrm{d}}{\mathrm{d}t}\mathrm{Exp}\left(\frac{Ht}{\mathrm{i}\hbar}\right)
    =H\ast\mathrm{Exp}\left(\frac{Ht}{\mathrm{i}\hbar}\right)}\nonumber\\
    &&~~=\left[\left(a_i\Big(x_{i}\!+\!\frac{\mathrm{i}\hbar}{2}\overrightarrow{\partial}\!_{p_i}
    \!+\!\frac{\mathrm{i}\mu}{2}\epsilon_{ij}\!\overrightarrow{\partial}\!_{x_j}\Big)
    \!+\!b_i\Big(p_{i}\!-\!\frac{\mathrm{i}\hbar}{2}\overrightarrow{\partial}\!_{x_i}
    \!+\!\frac{\mathrm{i}\nu}{2}\epsilon_{ij}\!\overrightarrow{\partial}\!_{p_j}\Big)\right)^2\right.\nonumber\\
    &&~~~~+\left.\!\left(c_i\Big(x_{i}\!+\!\frac{\mathrm{i}\hbar}{2}\overrightarrow{\partial}\!_{p_i}
    \!+\!\frac{\mathrm{i}\mu}{2}\epsilon_{ij}\!\overrightarrow{\partial}\!_{x_j}\Big)
    \!+\!d_i\Big(p_{i}\!-\!\frac{\mathrm{i}\hbar}{2}\overrightarrow{\partial}\!_{x_i}
    \!+\!\frac{\mathrm{i}\nu}{2}\epsilon_{ij}\!\overrightarrow{\partial}\!_{p_j}\Big)\right)^2\right]
    \mathrm{Exp}\left(\frac{Ht}{\mathrm{i}\hbar}\right).~~~~
\end{eqnarray}
After some straightforward algebras we arrive at the following form
\begin{equation}\label{expk}
    \mathrm{i}\hbar\frac{\mathrm{d}}{\mathrm{d}t}\mathrm{Exp}\left(\frac{Ht}{\mathrm{i}\hbar}\right)
    =\Big(H-k^2\partial_{\!_H}-k^2H\partial_{\!_H}^{2}\Big)\mathrm{Exp}\left(\frac{Ht}{\mathrm{i}\hbar}\right)~,
\end{equation}
where $H$ is the Hamiltonian~(\ref{hsq}), and
\begin{eqnarray}\label{kk}
k&=&(a_1d_1+a_2d_2-b_1c_1-b_2c_2)~\hbar+(a_1c_2-a_2c_1)~\mu+(b_1d_2-b_2d_1)~\nu\nonumber\\
&=&(\boldsymbol{a}\cdot\boldsymbol{d}-\boldsymbol{b}\cdot\boldsymbol{c})~\hbar
+(\boldsymbol{a}\wedge\boldsymbol{c})~\mu+(\boldsymbol{b}\wedge\boldsymbol{d})~\nu~.
\end{eqnarray}
Analogously, the generalized $\ast$-genvalue equation (\ref{eigen})
becomes
\begin{equation}\label{wigk}
    H\ast\mathcal{W}_{n}=\mathcal{W}_{n}\ast H=
    \Big(H-k^2\partial_{\!_H}-k^2H\partial_{\!_H}^2\Big)\mathcal{W}_{n}=E_{n}\mathcal{W}_{n}~.
\end{equation}
Obviously, Eqs. (\ref{expk}) and (\ref{wigk}) will work for any
value of the parameters $\mu$ and $\nu$, especially for $\mu=0$ and
$\nu=0$, which correspond to the ordinary quantum theory.

Now the Eqs. (\ref{expk}) and (\ref{wigk}) are much simpler than the
original form (\ref{time}) and (\ref{eigen}), they only contain one
variable $H$ in the right hand side of the equations. The solution
of Eq. (\ref{expk}) is \cite{s6} :
\begin{equation}\label{sexp}
   \mathrm{Exp}\left(\frac{Ht}{\mathrm{i}\hbar}\right)
   =\frac{1}{\cos\left(kt/\hbar\right)}\exp\left(\frac{H}{\mathrm{i}k}\tan\frac{kt}{\hbar}\right)~.
\end{equation}
Using the generating function for the Laguerre polynomials
\begin{equation}\label{lag}
    \frac{1}{1+s}\exp\left(\frac{zs}{1+s}\right)=\sum_{n=0}^{\infty}s^{n}(-1)^{n}L_{n}(z)~,
\end{equation}
where $L_{n}(z)$ is the Laguerre polynomial, with
$s=e^{-2\mathrm{i}kt/\hbar}$ and $z=2H/k$~, one can write the
$\ast$-Exponential function~(\ref{sexp}) as
\begin{equation}\label{lexp}
\mathrm{Exp}\left(\frac{Ht}{\mathrm{i}\hbar}\right)
=\sum_{n=0}^{\infty}2(-1)^{n}e^{-\mathrm{i}kt(2n+1)/\hbar}e^{-H/k}L_{n}\left(\frac{2H}{k}\right)~.
\end{equation}
Comparing this expression with the Fourier-Dirichlet expansion for
the time evolution function~(\ref{we}), we get the Wigner functions
and the corresponding energy eigenvalues\footnote{In this work we
will always ignore the normalized constant of any Wigner
functions.}:
\begin{equation}\label{wig}
    \mathcal{W}_{n}=e^{-H/k}L_{n}\left(\frac{2H}{k}\right)~,\qquad
    E_{n}=(2n+1)k~.
\end{equation}
They are also the solutions of Eq. (\ref{wigk}).

Obviously, for most physical systems their Hamiltonian functions are
not exactly the form~(\ref{hsq}), but we can use it to simplify our
calculation in some cases, for example in the case of coupled
harmonic oscillators.

\section{Coupled harmonic oscillators on noncommutative space}\label{sec4}
Now consider the two coupled harmonic oscillators
system \cite{s16,s17} on the phase space, the Hamiltonian can be
written as
\begin{equation}\label{H0}
H_0 = \frac{1}{2m_{1}}P^{2}_{1} + \frac{1}{2m_{2}} P^{2}_{2} +
\frac{1}{2} \left( C_1X^{2}_{1} + C_2 X^{2}_{2} + C_3 X_{1}
X_{2}\right),
\end{equation}
where $m_1$, $m_2$ are masses, and $C_1, C_2, C_3$ are constant
parameters. After rescaling the coordinates of the phase space
\begin{equation}\label{scale}
\begin{split}
x_{1}&= \left({\frac{m_{1}}{m_{2}}}\right)^{\frac{1}{4}} X_{1},
\quad x_{2} = \left({\frac{m_{2}}{m_{1}}}\right)^{\frac{1}{4}}
X_{2},\\
~p_{1}&= \left({\frac{m_{2}}{m_{1}}}\right)^{\frac{1}{4}} P_{1}~,
\quad p_{2} = \left({\frac{m_{1}}{m_{2}}}\right)^{\frac{1}{4}}
P_{2}~,
\end{split}
\end{equation}
one can rewrite $H_0$ as
\begin{equation}\label{H1}
H_1 = \frac{1}{2m}\left(p^{2}_{1} + p^{2}_{2} \right) +
\frac{1}{2}\left( c_1 x_{1}^{2} + c_2 x^{2}_{2} + c_3 x_{1} x_{2}
\right),
\end{equation}
where $m$, $c_1$, $c_2$ and $c_3$ are
\begin{equation}\label{para}
m = \sqrt{m_{1}m_{2}},\qquad c_1=C_1\sqrt{\frac{m_2}{m_1}}, \qquad
c_2=C_2\sqrt{\frac{m_1}{m_2}},\qquad c_3=C_3~.
\end{equation}
To remove the interaction term in $H_1$~(\ref{H1}), one may use the
following transformation
\begin{equation}\label{trans}
\begin{pmatrix}
  y_1 \\
  y_2 \\
\end{pmatrix}
=
\begin{pmatrix}
  \cos \frac{\alpha }{2} & -\sin \frac{\alpha }{2} \\
  \sin \frac{\alpha }{2} & ~\cos \frac{\alpha }{2} \\
\end{pmatrix}
\begin{pmatrix}
  x_1 \\
  x_2 \\
\end{pmatrix}
, \quad
\begin{pmatrix}
  q_1 \\
  q_2 \\
\end{pmatrix}
=
\begin{pmatrix}
  \cos \frac{\alpha }{2} & -\sin \frac{\alpha }{2} \\
  \sin \frac{\alpha }{2} & ~\cos \frac{\alpha }{2} \\
\end{pmatrix}
\begin{pmatrix}
  p_1 \\
  p_2 \\
\end{pmatrix},
\end{equation}
which is a unitary rotation with the mixing angle $\alpha$ on the
phase space. Applying the transformation~(\ref{trans}) to
(\ref{H1}), and letting the parameter $\alpha$ satisfy the condition
\begin{equation}\label{acon}
\tan\alpha  = \frac{c_3}{c_2 - c_1}~,
\end{equation}
one can find the Hamiltonian~(\ref{H1}) will be
\begin{equation}
\label{H2} H_2 = \frac{1}{2m} \left(q^{2}_{1} + q^{2}_{2} \right) +
\frac{K}{2}\left(e^{2\eta } y^{2}_{1} + e^{-2\eta }
y^{2}_{2}\right),
\end{equation}
where
\begin{equation}\label{keta}
K = \frac{1}{2}\sqrt{4c_1c_2 - c_3^{2}}~, \qquad  e^{2\eta}= \frac
{c_1 + c_2 + \sqrt{(c_1 - c_2)^{2} + c_3^{2}}}{\sqrt{4c_1c_2 -
c_3^{2}}}~,
\end{equation}
and the condition $4c_1c_2 > c_3^{2}$ must be fulfilled. Since the
transformation (\ref{trans}) does not change the structure of the
generalized $\ast$--product (\ref{gstar}), the coordinates $y_1$,
$y_2$, $q_1$ and $q_2$ should satisfy the commutation relations
(\ref{mbnc}) the same as those of $X_1$, $X_2$, $P_1$ and $P_2$~.

The Hamiltonian $H_{2}$~(\ref{H2}) can be separated into the
following two parts
\begin{equation}\label{hnc1}
\mathcal{H}_1= \left(\frac{e^{\eta}\sqrt{K}\sin
a}{\sqrt{2}}~y_1+\frac{\cos a}{\sqrt{2m}}~q_2\right)^2
+\left(\frac{e^{-\eta}\sqrt{K}\sin b}{\sqrt{2}}~y_2+\frac{\cos
b}{\sqrt{2m}}~q_1\right)^2~~,
\end{equation}
\begin{equation}\label{hnc2}
\mathcal{H}_2= \left(\frac{e^{\eta}\sqrt{K}\cos
a}{\sqrt{2}}~y_1-\frac{\sin a}{\sqrt{2m}}~q_2\right)^2
+\left(\frac{e^{-\eta}\sqrt{K}\cos b}{\sqrt{2}}~y_2-\frac{\sin
b}{\sqrt{2m}}~q_1\right)^2~.
\end{equation}
Obviously, with any value of $a$ and $b$, there should be
$\mathcal{H}_1+\mathcal{H}_2=H_2$.

Using the observation obtained in Sec.~\ref{sec3}~, we will
immediately get the expressions of the time-evolution functions for
$\mathcal{H}_1$~(\ref{hnc1}) and $\mathcal{H}_2$~(\ref{hnc2})
\begin{equation}\label{sexp1}
   \mathrm{Exp}_{1}\left(\frac{\mathcal{H}_1t}{\mathrm{i}\hbar}\right)
   =\frac{1}{\cos\left(k_{1}t/\hbar\right)}\exp\left(\frac{\mathcal{H}_1}{\mathrm{i}k_{1}}
   \tan\frac{k_{1}t}{\hbar}\right)~,
\end{equation}
\begin{equation}\label{sexp2}
   \mathrm{Exp}_{2}\left(\frac{\mathcal{H}_2t}{\mathrm{i}\hbar}\right)
   =\frac{1}{\cos\left(k_{2}t/\hbar\right)}\exp\left(\frac{\mathcal{H}_2}{\mathrm{i}k_{2}}
   \tan\frac{k_{2}t}{\hbar}\right)~,
\end{equation}
where
\begin{equation}\label{k1k2}
\begin{split}
    k_1&=\frac{\hbar\sqrt{K}}{2\sqrt{m}}\left(e^{\eta}\sin a\cos b-e^{-\eta}\sin b\cos
    a\right)+\frac{K\mu}{2}\sin a\sin b-\frac{\nu}{2m}\cos a\cos
    b~,\\
    k_2&=\frac{\hbar\sqrt{K}}{2\sqrt{m}}\left(e^{-\eta}\sin a\cos b-e^{\eta}\sin b\cos
    a\right)+\frac{K\mu}{2}\cos a\cos b-\frac{\nu}{2m}\sin a\sin
    b~.
\end{split}
\end{equation}
The Wigner functions and the corresponding energy spectra are
\begin{equation}\label{wig1}
    \mathcal{W}^{^{(1)}}_{n_1}=e^{-\mathcal{H}_1/k_1}L_{n_1}\!\left(\frac{2\mathcal{H}_1}{k_1}\right)~,\quad
    E^{^{(1)}}_{n_1}=(2n_1+1)k_1~;
\end{equation}
\begin{equation}\label{wig2}
    \mathcal{W}^{^{(2)}}_{n_2}=e^{-\mathcal{H}_2/k_2}L_{n_2}\!\left(\frac{2\mathcal{H}_2}{k_2}\right)~,\quad
    E^{^{(2)}}_{n_2}=(2n_2+1)k_2~.
\end{equation}
It is easy to verify when $a$ and $b$ take the following values
\begin{equation}\label{aspb}
\begin{split}
    \sin(a-b)&=\frac{\hbar\sqrt{Km}(e^{\eta}+e^{-\eta})}{\beta_1}~,\quad
    \cos(a-b)=\frac{Km\mu-\nu}{\beta_1}~,\\
    \sin(a+b)&=\frac{\hbar\sqrt{Km}(e^{\eta}-e^{-\eta})}{\beta_2}~,\quad
    \cos(a+b)=-\frac{Km\mu+\nu}{\beta_2}~,
\end{split}
\end{equation}
or
\begin{equation}\label{abtan}
\begin{split}
    a&=\frac{1}{2}\left(\arctan\frac{\hbar\sqrt{Km}(e^{\eta}-e^{-\eta})}{-(Km\mu+\nu)}
    +\arctan\frac{\hbar\sqrt{Km}(e^{\eta}+e^{-\eta})}{Km\mu-\nu}\right)~,\\
    b&=\frac{1}{2}\left(\arctan\frac{\hbar\sqrt{Km}(e^{\eta}-e^{-\eta})}{-(Km\mu+\nu)}
    -\arctan\frac{\hbar\sqrt{Km}(e^{\eta}+e^{-\eta})}{Km\mu-\nu}\right)~,
\end{split}
\end{equation}
then $\mathcal{H}_1$~(\ref{hnc1}) will be commutative with
$\mathcal{H}_2$~(\ref{hnc2}) under the Moyal bracket
\begin{equation}\label{hhc}
    [\mathcal{H}_1,\mathcal{H}_2]_{\ast}=\mathcal{H}_1\ast\mathcal{H}_2-\mathcal{H}_2\ast\mathcal{H}_1=0~,
\end{equation}
and their generalized $\ast$--product is equal to their ordinary
product
\begin{equation}\label{hheq}
    \mathcal{H}_1\ast\mathcal{H}_2=\mathcal{H}_1\mathcal{H}_2=\mathcal{H}_2\ast\mathcal{H}_1~.
\end{equation}
$\beta_1$ and $\beta_2$ in (\ref{aspb}) are
\begin{equation}\label{beta}
\begin{split}
    \beta_1&=\sqrt{(e^{\eta}+e^{-\eta})^{2}\hbar^2Km+(Km\mu-\nu)^2}~,\\
    \beta_2&=\sqrt{(e^{\eta}-e^{-\eta})^{2}\hbar^2Km+(Km\mu+\nu)^2}~,
\end{split}
\end{equation}
or
\begin{equation}\label{beta012}
    \beta_1=(e^{\eta}+e^{-\eta})\hbar\sqrt{Km}\sqrt{1+\Delta_1}~,\quad
    \beta_2=(e^{\eta}-e^{-\eta})\hbar\sqrt{Km}\sqrt{1+\Delta_2}~,
\end{equation}
with
\begin{equation}\label{delta12}
    \Delta_1=\frac{(Km\mu-\nu)^2}{(e^{\eta}+e^{-\eta})^{2}\hbar^2Km}~,
    \quad\Delta_2=\frac{(Km\mu+\nu)^2}{(e^{\eta}-e^{-\eta})^{2}\hbar^2Km}~,
\end{equation}
here $\Delta_1$ and $\Delta_2$ denote the effect of the
noncommutativity of the phase space. When $\mu=0$ and $\nu=0$~,
$\Delta_1=\Delta_2=0$~, it returns to the ordinary commutative phase
space.

With (\ref{k1k2}) and (\ref{aspb}), $k_1$ and $k_2$ can be written
as
\begin{equation}\label{k0102}
\begin{split}
    k_1&=\frac{1}{4m}(\beta_1+\beta_2)
    =\frac{\hbar\omega}{4}\left((e^{\eta}+e^{-\eta})\sqrt{1+\Delta_1}~+(e^{\eta}-e^{-\eta})\sqrt{1+\Delta_2}~\right)~,\\
    k_2&=\frac{1}{4m}(\beta_1-\beta_2)
    =\frac{\hbar\omega}{4}\left((e^{\eta}+e^{-\eta})\sqrt{1+\Delta_1}~-(e^{\eta}-e^{-\eta})\sqrt{1+\Delta_2}~\right)~,
\end{split}
\end{equation}
where $\omega=\sqrt{K/m}$~. Then from the definition of the
$\ast$-Exponential function~(\ref{expd}) and Eq.~(\ref{hheq}), we
obtain the time evolution function for the coupled harmonic
oscillators $H_{2}$~(\ref{H2}) on the noncommutative phase space
\begin{eqnarray}\label{exph}
\lefteqn{\mathrm{Exp}_{H_{2}}\left(\frac{H_{2}t}{\mathrm{i}\hbar}\right)
    =\mathrm{Exp}_{1}\left(\frac{\mathcal{H}_1t}{\mathrm{i}\hbar}\right)\ast
    \mathrm{Exp}_{2}\left(\frac{\mathcal{H}_2t}{\mathrm{i}\hbar}\right)}\nonumber\\
    &&~~~~~~~~~~~~=\mathrm{Exp}_{1}\left(\frac{\mathcal{H}_1t}{\mathrm{i}\hbar}\right)
    \mathrm{Exp}_{2}\left(\frac{\mathcal{H}_2t}{\mathrm{i}\hbar}\right)\nonumber\\
    &&~~~~~~~~~~~~=\frac{1}{\cos\left(k_{1}t/\hbar\right)}\exp\left(\frac{\mathcal{H}_1}{\mathrm{i}k_{1}}
    \tan\frac{k_{1}t}{\hbar}\right)
    \frac{1}{\cos\left(k_{2}t/\hbar\right)}\exp\left(\frac{\mathcal{H}_2}{\mathrm{i}k_{2}}
    \tan\frac{k_{2}t}{\hbar}\right),~~~~
\end{eqnarray}
and the Wigner functions are
\begin{eqnarray}\label{wigh}
    \mathcal{W}_{n_{1}n_{2}}&=&\mathcal{W}^{^{(1)}}_{n_1}\ast\mathcal{W}^{^{(2)}}_{n_2}
    =\mathcal{W}^{^{(1)}}_{n_1}\mathcal{W}^{^{(2)}}_{n_2}\nonumber\\
    &=&e^{-\mathcal{H}_1/k_1-\mathcal{H}_2/k_2}L_{n_1}\left(\frac{2\mathcal{H}_1}{k_1}\right)
    L_{n_2}\left(\frac{2\mathcal{H}_2}{k_2}\right)~,
\end{eqnarray}
the corresponding energies are
\begin{eqnarray}\label{enh}
E_{n_1n_2}&=&E^{^{(1)}}_{n_1}+E^{^{(2)}}_{n_2}=(2n_1+1)k_1+(2n_2+1)k_2\nonumber\\
    &=&\frac{1}{2m}\Big((n_1+n_2+1)\beta_1+(n_1-n_2)\beta_2\Big)\nonumber\\
    &=&\frac{\hbar\omega}{2}\Big((n_1+n_2+1)(e^{\eta}+e^{-\eta})\sqrt{1+\Delta_1}~+\nonumber\\
    &&~~~~~~+(n_1-n_2)(e^{\eta}-e^{-\eta})\sqrt{1+\Delta_2}~\Big)~.
\end{eqnarray}

Through the inverse transform of (\ref{trans}), all the results
above can be easily expressed in terms of the original variables of
the noncommutative phase space $(x_1,x_2,p_1,p_2)$ or
$(X_1,X_2,P_1,P_2)$~. So, based on the observation in
Sec.~\ref{sec3}, we derive a satisfactory result of the two coupled
harmonic oscillators system on the noncommutative phase space with
deformation quantization method, obtain all the Wigner functions and
the energy spectra.

\section{Some particular cases}\label{sec5}
\subsection{$\mu=0$ and $\nu=0$}
If $\mu=\nu=0$, then $\Delta_1=\Delta_2=0$, the noncommutative phase
space will reduce to the ordinary phase space. In this case, one has
\begin{equation}\label{2beta}
    \beta_1=(e^{\eta}+e^{-\eta})\hbar\sqrt{Km}~,\qquad\beta_2=(e^{\eta}-e^{-\eta})\hbar\sqrt{Km}~,
\end{equation}
and
\begin{equation}\label{k12}
    k_1=\frac{\hbar\sqrt{K}}{2\sqrt{m}}e^{\eta}=\frac{\hbar\omega}{2}e^{\eta}~~,\quad
    k_2=\frac{\hbar\sqrt{K}}{2\sqrt{m}}e^{-\eta}=\frac{\hbar\omega}{2}e^{-\eta}~.
\end{equation}
From (\ref{aspb}) we have
\begin{equation}\label{ab1}
    \sin(a-b)=1~,~\cos(a-b)=0~,\quad
    \sin(a+b)=1~,~\cos(a+b)=0~,
\end{equation}
so we may choose $a=\pi/2$ and $b=0$. Then
$\mathcal{H}_1$~(\ref{hnc1}) and $\mathcal{H}_2$~(\ref{hnc2}) become
\begin{equation}\label{HH12}
\mathcal{H}_1=\frac{1}{2m}q^{2}_{1}+ \frac{K}{2}e^{2\eta }
y^{2}_{1}~~,\quad
\mathcal{H}_2=\frac{1}{2m}q^{2}_{2}+\frac{K}{2}e^{-2\eta }
y^{2}_{2}~,
\end{equation}
and the Wigner functions are
\begin{eqnarray}\label{wigh1}
\mathcal{W}_{n_1n_2}&=&e^{-\mathcal{H}_1/k_1-\mathcal{H}_2/k_2}L_{n_1}
    \left(\frac{2\mathcal{H}_1}{k_1}\right)L_{n_2}\left(\frac{2\mathcal{H}_2}{k_2}\right)\nonumber\\
    &=&\exp\left\{-\frac{1}{\hbar\sqrt{Km}}\big(e^{-\eta}q_1^2+e^{\eta}q_2^2\big)-\frac{\sqrt{Km}}{\hbar}
    \left(e^{\eta}y_1^2+e^{-\eta}y_2^2\right)\right\}\nonumber\\
    &&~\times~L_{n_1}\!\left(\frac{2}{\hbar\sqrt{Km}}e^{-\eta}q_1^2+\frac{2\sqrt{Km}}{\hbar}e^{\eta}y_1^2\right)\nonumber\\
    &&~\times~L_{n_2}\!\left(\frac{2}{\hbar\sqrt{Km}}e^{\eta}q_2^2+\frac{2\sqrt{Km}}{\hbar}e^{-\eta}y_2^2\right),~~~~
\end{eqnarray}
with the corresponding energy spactra
\begin{eqnarray}\label{enh1}
    E_{n_1n_2}&=&(2n_1+1)k_1+(2n_2+1)k_2
    =\frac{\hbar\sqrt{K}}{2\sqrt{m}}\Big(e^{\eta}(2n_1+1)+e^{-\eta}(2n_2+1)\Big)\nonumber\\
    &=&\hbar\omega\left(e^{\eta}\Big(n_1+\frac{1}{2}\Big)+e^{-\eta}\Big(n_2+\frac{1}{2}\Big)\right)~.
\end{eqnarray}
These results are exactly the same as those in Ref.~\cite{s17}~.

\subsection{$\mu=-\nu=\theta$}
For simplicity, we also choose $e^{\eta}=K=m=1$ (e.g. let
$c_1$=$c_2$=$1$ and $c_3$=$0$ in (\ref{keta})), then $\beta_1$,
$\beta_2$ and $k_1$, $k_2$ become
\begin{equation}\label{beta3}
\begin{split}
    \beta_1&=2\sqrt{\hbar^2+\theta^2}~,\quad\beta_2=0~;\\
    k_1&=k_2=\frac{\sqrt{\hbar^2+\theta^2}}{2}~,
\end{split}
\end{equation}
and
\begin{equation}\label{sinasb}
    \sin(a-b)=\frac{\hbar}{\sqrt{\hbar^2+\theta^2}}~,\quad\cos(a-b)=\frac{\theta}{\sqrt{\hbar^2+\theta^2}}~,
\end{equation}
but the value $(a+b)$ can not be fixed, so we may choose $a=\pi$ for
simplicity, then we have
\begin{equation}\label{h1h23}
    \mathcal{H}_1=\frac{1}{2}\left(\frac{(\hbar y_2-\theta
    q_1)^2}{\hbar^2+\theta^2}+q_2^2\right)~,\quad
    \mathcal{H}_2=\frac{1}{2}\left(y_1^2+\frac{(\hbar q_1+\theta
    y_2)^2}{\hbar^2+\theta^2}\right)~.
\end{equation}
Thus the Wigner functions are
\begin{eqnarray}\label{wigh2}
\mathcal{W}_{n_1n_2}&=&e^{-\mathcal{H}_1/k_1-\mathcal{H}_2/k_2}L_{n_1}
    \left(\frac{2\mathcal{H}_1}{k_1}\right)L_{n_2}\left(\frac{2\mathcal{H}_2}{k_2}\right)\nonumber\\
    &=&\exp\left\{-\frac{1}{\sqrt{\hbar^2+\theta^2}}\left(y_1^2+\frac{(\hbar q_1+\theta
    y_2)^2}{\hbar^2+\theta^2}+\frac{(\hbar y_2-\theta
    q_1)^2}{\hbar^2+\theta^2}+q_2^2\right)\right\}\nonumber\\
    &&\times~L_{n_1}\left[\frac{2}{\sqrt{\hbar^2+\theta^2}}
    \left(\frac{(\hbar y_2-\theta q_1)^2}{\hbar^2+\theta^2}+q_2^2\right)\right]\nonumber\\
    &&\times~L_{n_2}\left[\frac{2}{\sqrt{\hbar^2+\theta^2}}
    \left(y_1^2+\frac{(\hbar q_1+\theta
    y_2)^2}{\hbar^2+\theta^2}\right)\right]~,
\end{eqnarray}
and the corresponding energies are
\begin{eqnarray}\label{enh2}
    E_{n_1n_2}&=&\frac{1}{2}\Big((n_1+n_2+1)\beta_1+(n_1-n_2)\beta_2\Big)\nonumber\\
    &=&(n_1+n_2+1)\sqrt{\hbar^2+\theta^2}~.
\end{eqnarray}
Therefore, in the case of $\mu=-\nu=\theta$, (\ref{wigh}) and
(\ref{enh}) will reduce to the results of Ref.~\cite{s13}.

\subsection{$\mu,\nu\ll\hbar$}
When $\mu,\nu\ll\hbar$~, it leads to $\Delta_1,\Delta_2\ll 1$, and
\begin{equation}\label{sqde}
    \sqrt{1+\Delta_1}\approx 1+\frac{\Delta_1}{2}~,
    \qquad\sqrt{1+\Delta_2}\approx 1+\frac{\Delta_2}{2}~.
\end{equation}
Thus the energy~(\ref{enh}) becomes
\begin{eqnarray}\label{en3}
    E_{n_1n_2}&=&\frac{\hbar\omega}{2}\Big((n_1+n_2+1)(e^{\eta}+e^{-\eta})\sqrt{1+\Delta_1}~
    +(n_1-n_2)(e^{\eta}-e^{-\eta})\sqrt{1+\Delta_2}~\Big)\nonumber\\
    &\approx&\hbar\omega\left[e^{\eta}\Big(n_1+\frac{1}{2}\Big)+e^{-\eta}\Big(n_2+\frac{1}{2}\Big)+\right.\nonumber\\
    &&~~~~~+\left.\frac{n_1+n_2+1}{4}(e^{\eta}+e^{-\eta})\Delta_1+\frac{n_1-n_2}{4}(e^{\eta}-e^{-\eta})\Delta_2\right].
\end{eqnarray}
Comparing this result with the energy spectrum~(\ref{enh1}) on the
commutative phase space, we find that, if the noncommutativity
parameters $\mu$ and $\nu$ are much smaller than the Planck
constant, the noncommutativity of the space structure will cause a
shift of the energy spectrum of the coupled harmonic oscillator
system. So if we find a deviation of the spectra of the coupled
harmonic oscillators system from the standard ones in some
experiments, we may measure and determine the values of the
noncommutativity parameters $\mu$ and $\nu$ and investigate the
effects of the noncommutativity.

\section{Conclusion}\label{sec6}
In this paper, we first consider deformation quantization for a
class of systems with special Hamiltonian. Using this method, we
investigate the two coupled harmonic oscillators system on a general
noncommutative phase space, and obtain the explicit expression of
the time-evolution function for this system, and furthermore, derive
all the Wigner functions and the corresponding energies. Since the
study is for a very general noncommutative space, and the results
are also rigorous and exact, so the results certainly can be reduced
to specific expressions reported in the literature in different
specific cases. In particular, the results show that when the spatial
or the momentum coordinates do not commute with each other, the energy
spectra of the coupled harmonic oscillators system will have a shift
comparing to the standard ones, and this phenomenon indicates a
possible way to explore the noncommutativity of the spatial or the
momentum coordinates in future experiments. When both the spatial
and the momentum coordinates are commutative with each other, the
results return to those of the ordinary case. We also discuss a case
in which the noncommutativity parameters $\mu$ and $\nu$ are much
smaller than the Planck constant, and calculate the quantity of the
energy shift, so if we could compare the quantity with some relevant
experimental data, we would get the bound of the noncommutativity
parameters, which will be very significant.

All these results come from our observation in Sec.~\ref{sec3}, and
we believe that this method should be useful in some other cases.

\section*{Acknowledgments}
This project was supported by the National Natural Science Foundation of China under Grant
Nos. 10375056 and 10675106.

\end{document}